\documentclass[12pt]{article}
\usepackage{authblk}
\usepackage{geometry}
\usepackage[dvipdfmx]{graphicx} 
\usepackage{soul}
\usepackage{ulem}
\usepackage{setspace}
\title{\Huge Charge-Density Waves of Single and Double NbS$_{3}$ Chains}

\author[1, 2, 4]{S. Tanda}
\author[2]{S. Kashimoto}
\author[2]{H. Yamamoto}
\author[5]{K. Inagaki}
\author[3, 4]{H. Nobukane}
\author[2]{Y. Fukuda}
\affil[1]{Research Institute of Electronic Science, Hokkaido University\\ Kita 20 Nishi 10, Kita-ku, Sapporo-city, 001-0020 Japan}
\affil[2]{Department of Applied Physics, Hokkaido University\\ Kita 13 Nishi 8, Kita-ku, Sapporo-city, 060-8628 Japan}
\affil[3]{Department of Physics, Hokkaido University\\ Kita 10 Nishi 8, Kita-ku, Sapporo-city, 060-0810 Japan}
\affil[4]{Center of Education and Research for Quantum Topological Science and Technology, Hokkaido University\\ Kita 13 Nishi 8, Kita-ku, Sapporo-city, 060-8628 Japan}
\affil[5]{Asahikawa Medical University\\ Midorigaoka Higashi 2 Jo 1 Chome, Asahikawa-city 078-8510, Japan}

\date{}
\begin{document}
\maketitle
\section*{Abstract}
The physics of a genuine one-dimensional system in which electrons are confined in one direction remains unclear. The actual electronic state of such a genuinely one-dimensional system has not been investigated in previous experiments, for they have all been conducted on quasi-one-dimensional specimens, namely in strongly anisotropic bulk crystals \cite{Monceau}. Conventionally, charge-density waves (CDWs) driven by Fermi surface nesting have been considered to appear in one-dimensional electron-lattice systems \cite{Peierls, KagoshimaSambongi,Gruner}. However, the CDW transitions actually observed to date have all occurred in quasi-one-dimensional systems and therefore do not directly indicate a genuine one-dimensional electronic state. 

We investigated, for the first time, isolated single and double-chain NbS$_{3}$ samples using the carbon-nanotube-sheath method \cite{Pham} and discovered CDWs in both systems. In the single-chain, surprisingly, a $(1/4)b^*$ CDW was observed, in contrast to the $(1/3)b^*$ CDW that has been observed in bulk samples \cite{Boswell} . In the double-chain, the coexistence of a $(1/2)b^*$ dimer structure  and a $(1/3)b^*$ CDW was confirmed. This discovery represents a significant advancement in the field of low-dimensional physics, surpassing the limitations of previous studies on bulk systems.

\section{Introduction}

One-dimensional systems in which electrons are confined in one direction have been an attractive issue for decades. However, previous experiments have all been conducted on quasi-one-dimensional specimens, namely, in strongly anisotropic bulk crystals \cite{Monceau,Peierls, KagoshimaSambongi,Gruner}. Theoretically, the emergence of the Luttinger liquid has been expected in such systems\cite{Luttinger} . Indeed, some experimental results in carbon nanotubes have been interpreted as a Luttinger liquid, based on the power-law temperature dependence of electrical conductivity and photoemission intensity \cite{Bockrath, Ishii}. These systems are, however, not genuinely one-dimensional due to the presence of quantised degrees of freedom in the circumferential direction.

Conventionally, charge-density waves (CDWs) driven by Fermi-surface nesting have been considered to emerge in one-dimensional electron-lattice systems \cite{Peierls, KagoshimaSambongi,GrunerZettle}, whereas the CDW transitions actually observed occur in pseudo-one-dimensional systems and do not directly indicate a genuine one-dimensional electronic state. Indeed, in NbSe$_{3}$, one of the transition metal trichalcogenides (TMTs), two of the three crystallographically distinct Nb chains exhibit CDW transitions at approximately 145 K and 59 K, respectively, while the third chain keeps metallic even at low temperatures \cite{Ong}. The reason why the metallic state survives, rather than yields to a CDW, remains an open problem. To uncover the origin of this behaviour, it is essential to investigate an isolated single chain.

As an ideal candidate for investigating genuine one-dimensional systems, we focused on NbS$_{3}$ exhibiting a CDW transition at room temperature in bulk samples and expected it to form an isolated single chain. The crystal structure of NbS$_{3}$ consists of Nb atom chains arranged in a one-dimensional configuration, with Nb atoms positioned at the centres of triangular prisms formed by S atoms (Figure 1) \cite{Zettl}. In bulk samples, NbS$_{3}$ exhibits multiple polymorphs arising from variations in chain twisting, spatial arrangement, and interchain stacking patterns \cite{Conejeros}. These polymorphs give rise to distinct CDW behaviours, including differences in the presence, transition temperatures, and ordering of CDWs \cite{Zybtsev}. Such polymorphism complicates the understanding of CDW transitions, as the observed properties can vary significantly across different polymorphs. Recent advances have enabled the successful synthesis of single-chain TMT samples within carbon nanotube (CNT) sheaths, revealing their unique structural and physical properties \cite{Pham}. These samples exhibit novel phenomena, including static and dynamic torsional waves and charge transfer effects, pointing to the potential for further advances in the physics of genuine one-dimensional systems. The use of this fabrication technique to isolate NbS$_{3}$ chains removes the complexities associated with multiple polymorphs, allowing a clearer understanding of CDW formation and ordering in a genuine one-dimensional system.

We investigated isolated single- and double-chain NbS$_{3}$ samples using the CNT-sheath method and discovered CDWs in both systems. Scanning transmission electron microscopy (STEM), which enables atomic-scale observation, was used in this study to examine the Nb-Nb interatomic distances and their distribution within the Nb atomic chains. This approach revealed the characteristics of the resulting CDW order in isolated atomic chains. In the single-chain, a $(1/4)b^*$ CDW was observed, in contrast to the $(1/3)b^*$ CDW seen in bulk samples  \cite{Boswell}. In the double-chain, the coexistence of a dimer structure and a $(1/3)b^*$ CDW was confirmed. Notably, both of these phases represent charge-ordered states that have not been previously identified in bulk samples.

\section{Results}
Figure \ref{fig:STEMimages} shows room-temperature scanning transmission electron microscopy (STEM) images of (a) the single chain and (b) the double chain synthesised within CNTs. The white spots were determined to be individual Nb atoms, based on the difference between the atomic numbers of Nb and S,  as well as the actual length of the distance between the white spots in each row.
The exact atomic positions of the S and C atoms in CNTs could not be determined from these images because their corresponding contrasts were unclear. In this analysis, we attempted to determine the position of individual Nb atoms as \textit{precisely} as possible and to measure the distance between adjacent Nb atoms to extract CDW information.
We used the image processing software \cite{Rasband} for smoothing to reduce uncertainty in the atomic positions caused by limited resolution and noise  (Trimmed STEM images in Fig. \ref{fig:CDW}), and then deduced the atom centres from the determination of the local maxima of luminance in the images to estimate the distance between all neighbouring atoms.

\begin{figure}
    \centering
    \includegraphics[width=0.9\textwidth]{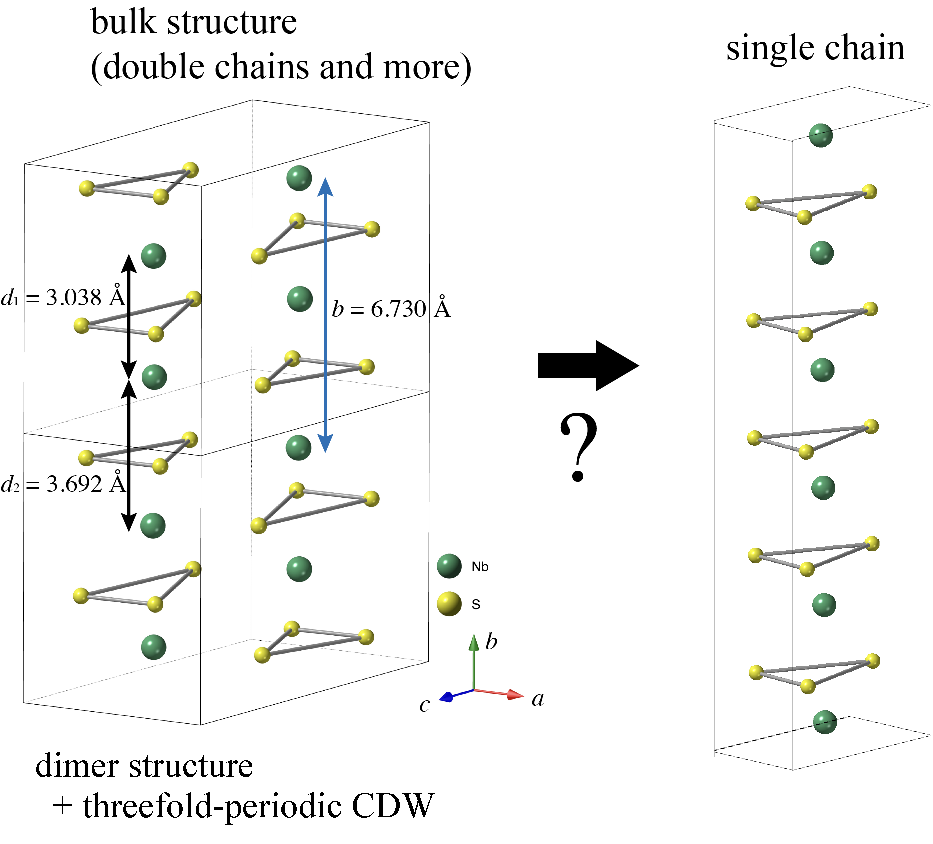}
    \caption{(Left) The crystal structure model of NbS$_{3}$-I \cite{Rijnsdorp}. The unit cell contains four Nb atoms and twelve S atoms. The Nb chains are aligned in the $b$-axis direction, exhibiting metallic electrical conductivity along this axis and forming a quasi-one-dimensional electron system. In this model, the Nb-Nb interatomic distance alternates between $d_{1}= 3.038$ \AA\ and $d_{2}= 3.692$ \AA, resulting in a doubly periodic dimer structure accompanied by threefold-periodic CDWs  \cite{Boswell, Rijnsdorp, Conejeros}. (Right) The behavior of an isolated, genuine one-dimensional electron system has remained unknown.}
    \label{fig:NbS3_long}
\end{figure}

\begin{figure}
    \centering
    \includegraphics[width=0.9\textwidth]{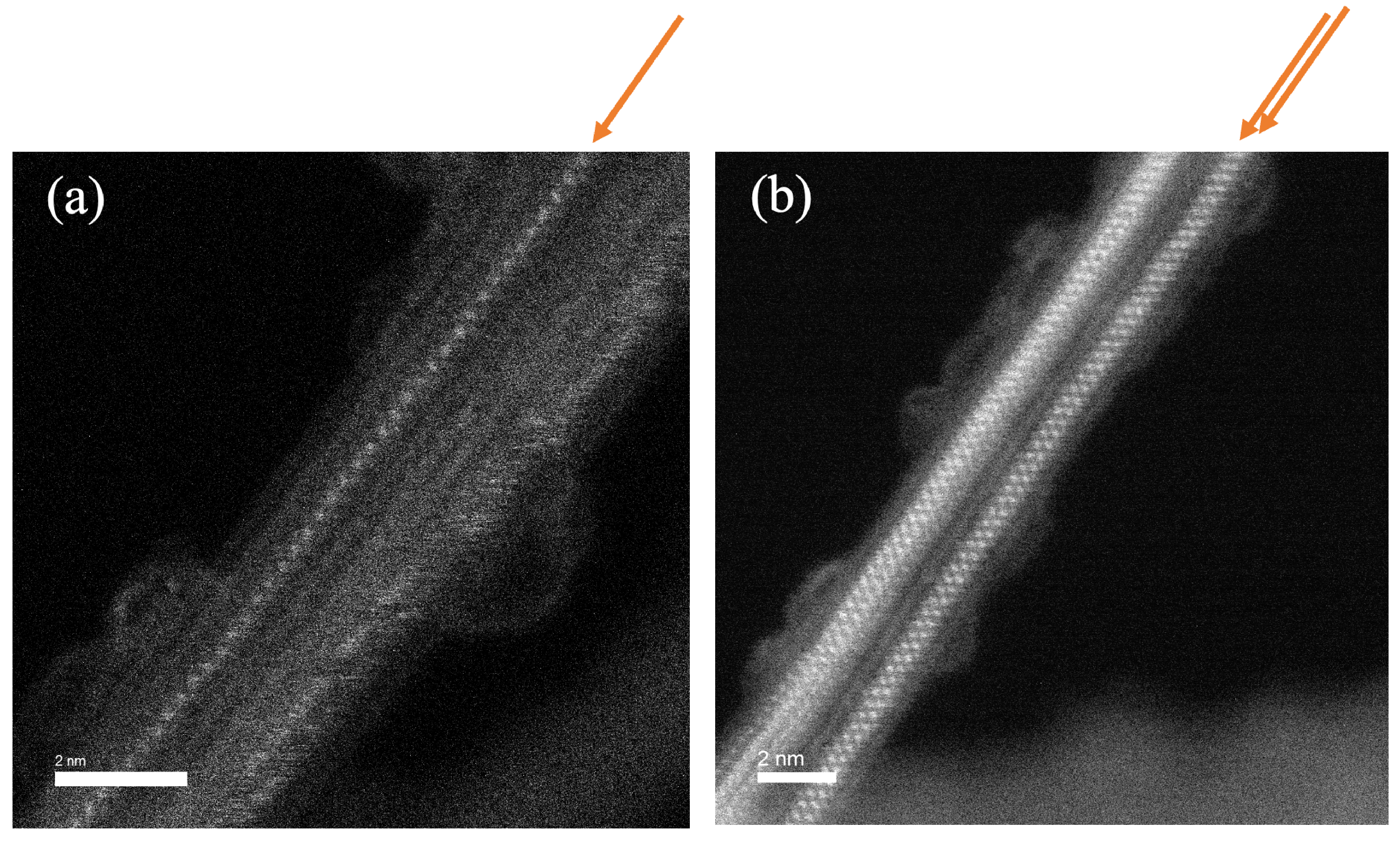}
    \caption{STEM images of NbS$_{3}$ synthesized in carbon nanotubes (CNTs). (a) The red arrow indicates an isolated single Nb atomic chain. Because the atomic number of Nb is larger than that of S, and C in CNT, the positions of individual Nb atoms can only be figured out. (b) The two red arrows indicate a double chain of Nb atoms. The direction in which the chains are aligned corresponds to the $b$-axis in the structural model (Fig. \ref{fig:NbS3_long}). In addition, there are three or more overlapping chains on the left side of the double chain.}
    \label{fig:STEMimages}
\end{figure}

\begin{figure}
    \centering
    \includegraphics[width=0.9\textwidth]{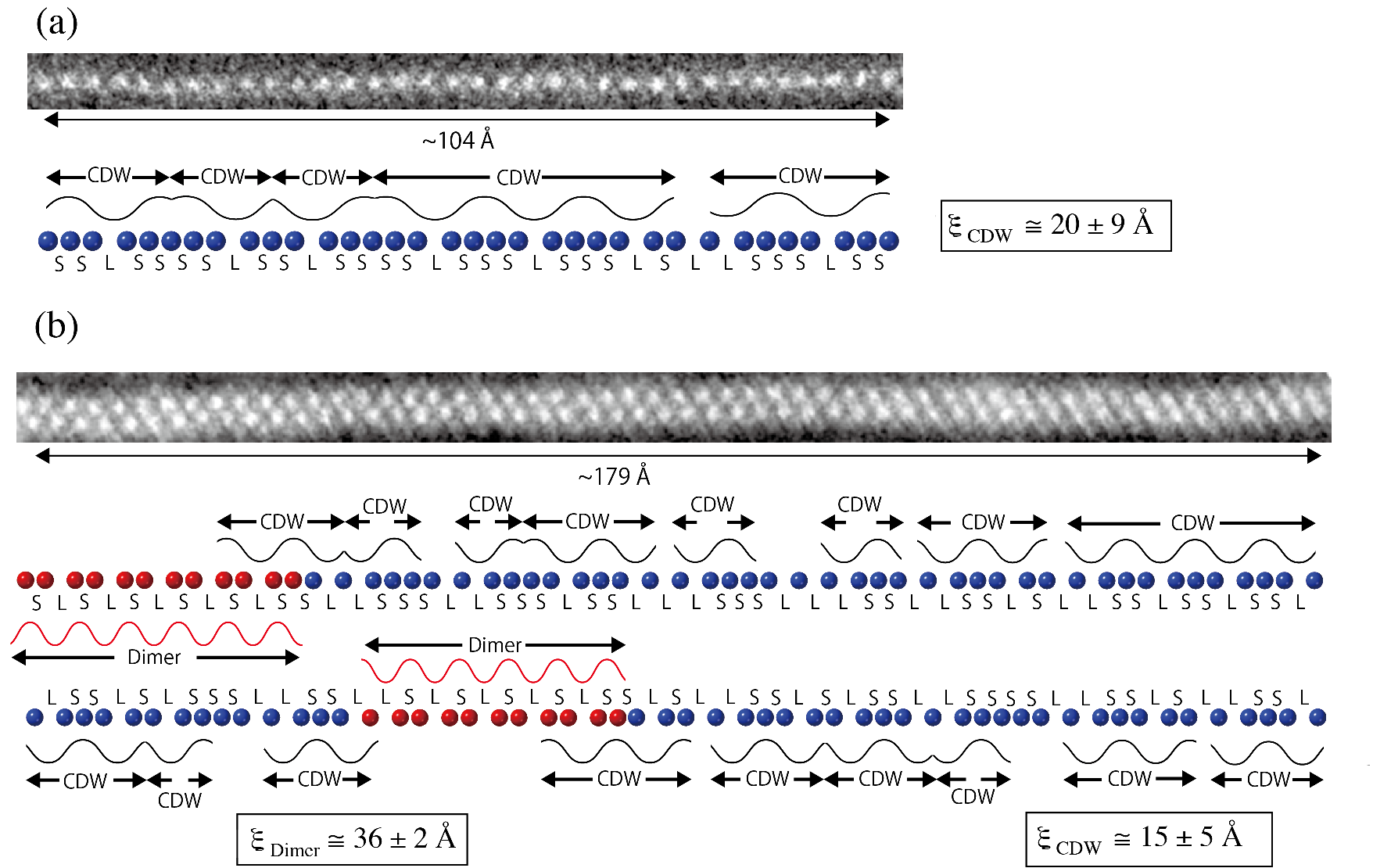}
    \caption{(a) Fourfold-periodic CDWs in the single chain, resulting from the position analysis in real space. The positions of the Nb atoms were read from the above STEM image using image processing (see text). The schematic illustration shows the distance between Nb-Nb atoms (the blue circles). These interatomic distances are categorised as either long (L) or short (S), in comparison to the average value 3.365 \AA\ for $d_{1}$ and $d_{2}$ (Fig. \ref{fig:NbS3_long}). \textit{No dimer structures are found in the single chain.} Instead, there are local $\cdots$(SLSS)$\cdots$ structures, as expected in a \textit{fourfold-periodic} CDW, with correlation length $\xi_\mathrm{CDW}=20\pm9$ \AA. The sinusoidal waves represent CDWs corresponding to Nb atomic displacements. Surprisingly, the average Nb-Nb interatomic distance was found to be approximately 3.15 \AA, shrunk by 6 \% of the bulk structure model (Fig. \ref{fig:NbS3_long}). (b) Coexistence of threefold-periodic CDWs and dimer structures in the double chain. As expected in threefold-periodic CDWs, the local $\cdots$(LSS)$\cdots$ structures occur with $\xi_\mathrm{CDW}=15\pm5$ \AA\ (the blue circles). Dimer structures appear in the form of $\cdots$(LS)(LS)(LS)$\cdots$ with $\xi_\mathrm{dimer}=36\pm 2$ \AA\ (the red circles). This coexistence is consistent with the studies of bulk samples as a quasi-one-dimensional electron system \cite{Boswell, Rijnsdorp, Conejeros}. The average Nb-Nb interatomic distance is approximately 3.38 \AA, same as that of bulk.}
    \label{fig:CDW}
\end{figure}

Figure \ref{fig:CDW} shows the results of the investigation of the distribution of interatomic distances obtained from the STEM images. The interatomic Nb-Nb distances obtained here are classified as L or S depending on whether they were longer or shorter than the average Nb-Nb distance $d_{Ave.} = 3.365$ \AA\ for $d_{1}$ and $d_{2}$ in Fig. \ref{fig:NbS3_long}. The blue and red circles in Fig. \ref{fig:CDW} represent Nb atoms, and the interatomic distances are schematically indicated by L and S.
In the lower part of Fig. \ref{fig:CDW}a, no sequence of reasonable length, such as $\cdots$(LS)(LS)(LS)$\cdots$, was observed;  instead, in all of the areas, S appeared more frequently than L.
This implies the existence of longer units, such as (LSS) and (SLSS).
The middle part of Fig. \ref{fig:CDW}a schematically shows sinusoidal waves representing CDWs corresponding the local sequence (SLSS). The correlation length of CDWs in this single chain was found to be $\xi_\mathrm{CDW}=20\pm9$ \AA. Another important result revealed in this study is that the average Nb-Nb distance in the single chain was found to be $\sim 3.156$ \AA, which \textit{shrank} by 6\% from that of the structural model (Fig. \ref{fig:NbS3_long}). These results confirm the occurrence of CDWs in the single chain of NbS$_{3}$ at room temperature. Neither a dimer structure nor threefold-periodic CDWs were found in contrast to those reported in bulk samples.

On the other hand, analysis of the double chain using the STEM image in Fig. \ref{fig:STEMimages}b yielded different results from the single chain. The dimer structures $\cdots$(LS)(LS)(LS)$\cdots$ partially appeared in both atomic chains (red circles). Although the site of appearance was shifted, the dimer structures in both chains had similar correlation length $\xi_\mathrm{dimer}=36\pm2$ \AA. In the other regions of the double chain, the threefold periodic (LSS) local structure was dominant, rather than the fourfold periodic (SLSS) structure found in the single chain. The correlation length of the threefold periodic CDWs in the two chains was determined to be $\xi_\mathrm{CDW}=15\pm5$ \AA. The average Nb-Nb interatomic distances were $\sim 3.377$ \AA\ and $\sim 3.383$ \AA\ for the two chains, respectively. These lengths are deviated by 0.3 \% and 0.5 \%, respectively, from the crystal structure model's value of 3.365 \AA. The coexistence of a dimer structure and threefold periodic CDW (LSS) sequences together with comparable average interatomic distances, confirms that the double chain exhibits properties similar to those of NbS$_{3}$ bulk, which is a quasi-one-dimensional electron system.

\section{Discussions}

The structure model of NbS$_{3}$-I type is characterised by alternating Nb-Nb interatomic distances along the b-axis, with $d_{1}= 3.038$ \AA\ and $d_{2}= 3.692$ \AA\ (Fig. \ref{fig:NbS3_long}). In this paper, the crystal structure containing this doubly periodic arrangement of Nb atoms is referred to as the ``dimer structure''. A threefold-periodic CDW has been known in NbS$_{3}$-II type since the early days of NbS$_{3}$ research, and its existence has been confirmed by superlattice reflections appearing near $(\pm1/3)b^{*}$ in the vicinity of the basic diffraction spots of the basic lattice in electron diffraction images \cite{Boswell, Zettl}. Subsequently, it has been revealed that multiple phase transitions of CDW occur at 150 K, 360 K, and 450–475 K in NbS$_{3}$-II \cite{Zybtsev}. In the polymorphic structure of the bulk state, not only the arrangement of Nb atoms but also that of S atoms changes, each exhibiting characteristic long-range order. Consequently, a variety of physical properties such as CDW arise, which complicates the essential characteristics of this material. 

\begin{figure}
    \centering
    \includegraphics[width=0.9\textwidth]{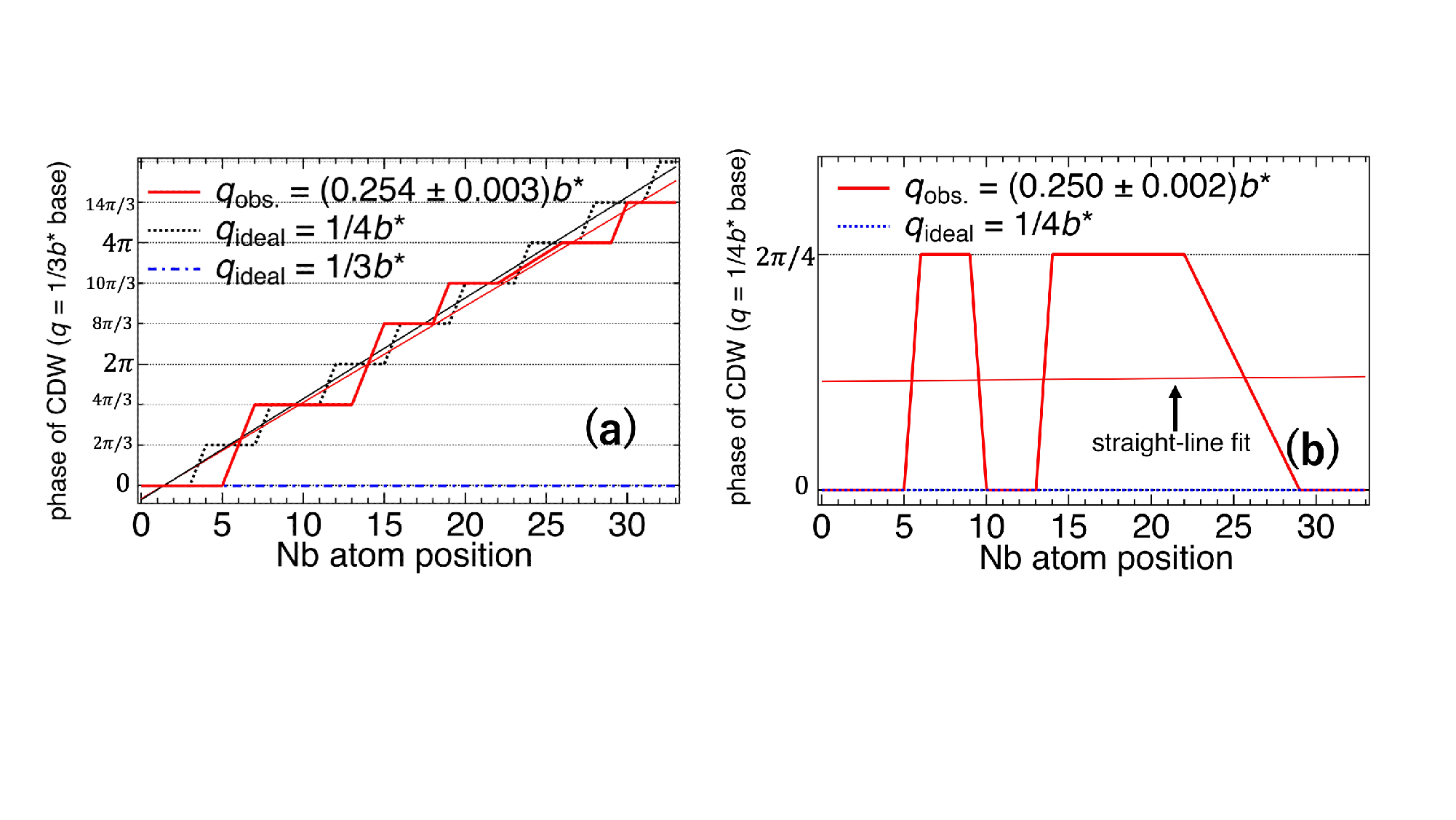}
    \caption{Phase analysis of CDW for the single chain. The horizontal axis represents the atomic position, while the vertical axis represents the phase shift caused by deviation from the reference commensurate wavenumber. (a) Results based on the threefold-periodic CDW. The red line shows the staircase-like phase shift whose step corresponds to a discommensuration between (LSS) units. The average slope of the stairecase represents the CDW wave number $q_{obs.}=(0.254 \pm 0.003)b$*. For a comparison, threefold and fourfold CDWs are shown as dotted lines. (b) Results based on fourfold-periodic CDW. In this analysis, the red line includes discomensurations with $\pm$ 2$\pi$/4 per atom displaced from $\cdots$(SLSS)(SLSS)(SLSS)$\cdots$. When local deviations from (SLSS) are negative, such as (SLSS)($SLS$)(SLSS) or (SLSS)($LSS$)(SLSS), where the phase is plotted with a 2$\pi$/4 reduction at the position. The blue dotted line shows the ideal $\cdots$(SLSS)(SLSS)(SLSS)$\cdots$ case, the phase shift is always zero. The thin red solid line is the result of a straight-line fit. The gradient obtained from the fit was almost zero, and $q_{obs.}$ was 0.25$b$*. This result suggests that fourfold-periodic CDWs occur in the single chain.}
    \label{fig:discomme}
\end{figure}

 The results obtained in the present study differ from those reported previously. We investigated whether the atomic configuration characteristic of the genuine single chain, as revealed by the present study, is the result of a particular CDW wavenumber. We then performed a phase analysis based on McMillan's \cite{McMillan} concept of discommensuration (DC), which treats DCs as defects in a commensurate CDW. Figure \ref{fig:discomme} shows the results of the analysis of the atomic arrangement of the single chain. We first performed this analysis using the commensurate wavenumber 1/3$b^{*}$ corresponding to the threefold unit (LSS). The horizontal axes represent the atomic positions, with all inter-atomic distances normalised to 1. The vertical axes show the phase shift at that position, which increases by the $b^{*}$ component of the reference commensurate wavenumber whenever atomic separation L or S deviates from the ideal sequence.
In Fig. \ref{fig:discomme}a, the red staircase-like solid line shows the results obtained for the single chain in this study, where the shift corresponds to the discommensuration. In this analysis based on threefold periodicity, the average slope is given by $2\pi(1/3-k)$, where the wavenumber of the CDW is $kb^{*}$. Using the slope value obtained by fitting, we obtained $k=0.254\pm0.003$, which is almost equal to 1/4. As reference, the black dotted line shows a fourfold-periodic commensurate structure, while the blue dash-dot line shows a threefold-periodic commensurate one. From these comparisons, it is clear that the single chain exhibits CDWs with a fourfold periodicity, rather than a threefold periodicity.

Figure \ref{fig:discomme}b shows the results of the phase analysis based on the fourfold-periodic (SLSS) local structure. In this case the phase shift is $2\pi(1/4)$ for local deviations from the ideal commensurate structure. These phase shifts due to deviation is plotted as soliton and anti-soliton for over- and undershift relative to the commensurate (SLSS)(SLSS)(SLSS) sequence. When local deviations from (SLSS) are negative, such as (SLSS)(SLS)(SLSS) or (SLSS)(LSS)(SLSS), the phase is plotted with a $2\pi(1/4)$ reduction at the position. On the other hand, when local deviations from (SLSS) are positive, such as (SLSS)(SSLSS)(SLSS), the phase is plotted with a $2\pi(1/4)$ increase at that position.
The thin red solid line is the result of a straight-line fit. The gradient obtained from the fit was almost zero, and the observed wavevector $q_{obs.}= (0.250 \pm 0.002)b^*$. This result suggests that a locked-in phase transition to a fourfold-periodic commensurate CDW occurs in the single chain, and that soliton-antisoliton pairs maintain coherence at room temperature.

The present analysis does not lead to long-range order due to the short correlation length of CDW, which is consistent with the lack of long-range order at finite temperatures in $genuine$ one-dimensional systems with short-range interactions \cite{Landau}. 
It is therefore important to recognize that, even in true one-dimensional systems where long-range order is absent, an underlying mechanism governing atomic arrangement may still exist. 
For example, quasiperiodic structures arising from the nesting effect associated with a CDW may unexpectedly emerge in genuine one-dimensional systems. In fact, the average lattice constant in the single chain was found to be approximately 3.156 \AA. The 6\% shrinkage seemed too large at a glance. However, recent theoretical study proposes that the formation of CDW can be attributed with a change of the lattice constant \cite{Nakatsugawa}. This will be an important perspective.   
\section{Conclusion}
In conclusion, by using single-chain and double-chain samples of isolated NbS$_{3}$ synthesized within CNTs, we have demonstrated that genuine one-dimensionality gives rise to a CDW state that differs from that of the bulk material. 
In the single-chain sample, a fourfold-periodic commensurate CDW accompanied by discommensurations and a 6\% shrinkage relative to the bulk structural model was observed. In contrast, in the double-chain sample, the dimer structure and threefold-periodic CDWs coexist in each chain.
These results indicate that the genuine one-dimensional electronic system of NbS$_{3}$ is not a Luttinger liquid but the CDW state driven by the Peierls transition, and that the change in dimensionality also leads to a change in the universality class. This discovery represents a significant advance in the field of low-dimensional physics, surpassing the limitations of previous studies on bulk systems, which are quasi-one-dimensional.
\\



\newpage


\clearpage


\section*{Methods}
\textbf{Sample preparation}

 We have chosen NbS$_{3}$, a material with relatively large electron-lattice interactions and which forms CDWs from high temperatures in the bulk, and investigated its behavior by controlling dimensionality as an experimental parameter. To fabricate a single chain or a double chain of Nb in NbS$_{3}$, we attempted to suppress the spread of crystal growth in the a- and c-axis directions by using carbon nanotubes of $2 \-- 3$ nm in diameter. The materials, carbon nanotubes, and high-purity Nb and S, were placed in a quartz tube that had been heated while being evacuated beforehand to remove impurities. The molar ratio of Nb to S was set to 1 : 3. The quartz tube with materials was kept in a low vacuum of $\sim 1$ Pa while a gas burner was used to remove adsorbed molecules, finally, the tube was sealed in the low vacuum state. Several quartz tubes with materials were prepared using this method, and each was annealed in an electric furnace at 590 ${}^\circ$C or 600 ${}^\circ$C for one week. After the annealing process, the quartz tubes were quenched in water. The quenching process prevents the remaining S that was not used to form NbS$_{3}$ from adhesion on the surface of the carbon nanotubes. Carbon nanotubes removed from quartz tubes were dispersed by sonication in ethanol solvent. The carbon nanotubes with ethanol solvent were aspirated with a dropper, placed on a microgrid, and observed by scanning transmission electron microscopy (STEM, JEOL: JEM-ARM200F, 80 kV).

\begin{thebibliography}{99}

\bibitem{Monceau}
Monceau, P. Electronic crystals: An experimental overview. \textit{Advances in Physics} \textbf{61}, 325–581 (2012).

\bibitem{Peierls}
Peierls, R. E. \textit{Quantum Theory of Solids} Ch. 5 (Clarendon Press, Oxford, 1955).

\bibitem{KagoshimaSambongi} S. Kagoshima, H. Nagasawa and T. Sambongi, \textit{One-Dimensiomal Conductors}, Springer-Verlag, Berlin, 1988 

\bibitem{Gruner}
Gr\"{u}ner, G. \textit{The dynamics of charge-density waves}. Reviews of Modern Physics \textbf{119}, 117–232 (1985).

\bibitem{Pham}
Pham, T. \textit{et al}. Torsional instability in the single-chain limit of a transition metal trichalcogenaide. \textit{Science} \textbf{361}. 263-266 (2018).

\bibitem{Boswell}
Boswell, F. W. \& Prodan, A. Peierls distortions in NbS$_{3}$ and NbSe$_{3}$. \textit{Physica} \textbf{99B}, 361-364 (1980).


\bibitem{Luttinger}
Luttinger, J. M. An exactly soluble model of a many-fermion system. \textit{J. Math. Phys.} \textbf{4} 1154-1162 (1963).

\bibitem{Bockrath}
Bockrath, M. \textit{et al}. Luttinger-liquid behaviour in carbon nanotubes. \textit{Nature} \textbf{397}, 598-601 (1999).

\bibitem{Ishii}
Ishii, H. \textit{et al}. Direct observation of Tomonaga-Luttinger-liquid state in carbon nanotubes at low temperatures. \textit{Nature} \textbf{426}, 540-544 (2003).

\bibitem{GrunerZettle}
Gr\"{u}ner, G., Zettle, A. \textit{Charge density wave conduction: A novel collective transport phenomenon in solids}. Reviews of Modern Physics \textbf{60}, 1129–1181 (1988).

\bibitem{Ong}
Ong, N.P. Transport studies near phase transitions in NbSe$_{3}$. \textit{Phys. Rev. B}\textbf{17}, 3243–3252 (1978).

\bibitem{Zettl}
Zettl, A., Jackson, C. M., Janossy, A. \& Gr\"{u}ner, G. Charge density wave transition and nonliner conductivity in NbS$_{3}$. \textit{Solid State Commun.} \textbf{43}, 345-347 (1982).

\bibitem{Rijnsdorp}
J. Rijnsdorp, J. \& Jellinek, F. J. The crystal structure of Niobium Trisulfide, NbS$_{3}$ \textit{solid state chem.} \textbf{25}, 325-328 (1978).

\bibitem{Conejeros}
Conejeros, S. \textit{et al}. Rich polymorphism of layered NbS$_{3}$. \textit{Chem. Mater.} \textbf{33}, 5449-5463 (2021).

\bibitem{Zybtsev}
Zybtsev, S. G. \textit{et al}. NbS$_{3}$: A unique quasi-one-dimensional conductor with three charge density wave transitions. \textit{Phys. Rev. B} \textbf{95}, 035110 (2017).

\bibitem{Rasband}
Rasband, W.S., ImageJ, U. S. National Institutes of Health, Bethesda, Maryland, USA, http://imagej.nih.gov/ij/, 1997-2012.

\bibitem{McMillan}
McMillan, W. L. Theory of discommensurations and the comensurate-incommensurate charge-density-wave phase transition. \textit{Phys. Rev. B} \textbf{14}, 1496-1502 (1976).

\bibitem{Landau} Landau, L. D. and Lifshitz, E. M. \textit{Statistical Physics}, Part 1, 3rd ed., Vol. 5 of Course of Theoretical Physics (Pergamon Press, Oxford, 1980).

\bibitem{Nakatsugawa}
Private communication with Dr. K. Nakatsugawa and Dr. T. Fujii on lattice constant and commensurability of CDW.





\end{thebibliography}
\end{document}